\documentclass[conference]{IEEEtran}
\usepackage[]{graphicx}
\usepackage{bm} % use mathit and mathrm together by \bm
\usepackage{ctable,amsmath}
\newcommand{\comments}[1]{}

\DeclareMathOperator*{\argmax}{arg\,max}

\renewcommand{\figurename}{Fig.}

\include{now_macro}

\begin{document}
%\NowFootNum

\title{Improved Accent Classification Combining Phonetic Vowels with Acoustic Features}
\author{\IEEEauthorblockN{Zhenhao Ge}
%\IEEEauthorblockA{School of Electrical and Computer Engineering, Purdue University\\
%465 Northwestern Ave., West Lafayette, Indiana, USA, 47907-2035\\
%Email: zge@purdue.edu, Telphone: (317) 457-9348}
%\IEEEauthorblockA{Interactive Intelligence Inc.\\
%7601 Interactive Way, Indianapolis, Indiana, 46278\\
%Email: roger.ge@inin.com, Telphone: +1 (317) 457-9348}
\IEEEauthorblockA{School of Electrical and Computer Engineering \\
Purdue University, West Lafayette, Indiana, 47907, USA}
}
  
\maketitle

\begin{abstract}
Researches have shown accent classification can be improved by integrating semantic information into pure acoustic approach. In this work, we combine phonetic knowledge, such as vowels, with enhanced acoustic features to build an improved accent classification system. The classifier is based on Gaussian Mixture Model-Universal Background Model (GMM-UBM), with normalized Perceptual Linear Predictive (PLP) features. The features are further optimized by Principle Component Analysis (PCA) and Hetroscedastic Linear Discriminant Analysis (HLDA). Using 7 major types of accented speech from the Foreign Accented English (FAE) corpus, the system achieves classification accuracy 54\% with input test data as short as 20 seconds, which is competitive to the state of the art in this field.
\end{abstract}


\section{Introduction}
\label{sec:introduction}

As businesses become more international, accent verification and classification gains more attention recently, probably because of the increasing demand for better recognizing non-native speakers and their accented speech. However, this problem is still very challenging, since there are many types of accents and the response time allowed for accent detection is usually very short. Choueiter et al. achieved accuracy of 32\% classifying 23 types of accented English \cite{choueiter2008empirical}, using methods in language identification (LID), such as Maximum Mutual Information (MMI) training and Gaussian tokenization. Omar et al. recently integrated Universal Background Model (UBM) into Support Vector Machine (SVM) classifier and claimed that it outperformed the results in \cite{choueiter2008empirical} by 75.3\% relatively \cite{omar2010novel}. Another work for German vs. Spanish classification in \cite{macias2003acoustic} reported classification rates of 73\% and 58.9\%, using GMMs and naive Bayes classification respectively. In addition, classification rates of 36.2\%, 17.7\% and 13.2\% were reported in \cite{macias2003acoustic}, for 4-, 13- and 23-way classification using naive Bayes. To the best of our knowledge, these are the only three works, which used the same dataset used in this work.

In this work, we first created a baseline accent classifier for 7 selected types of English accents, using Gaussian Mixture Model-Universal Background Model (GMM-UBM), with normalized Perceptual Linear Predictive (PLP) features. The feature were then dimension-reduced and discriminatively optimized using Principle Component Analysis (PCA) and Heteroscedastic Linear Discriminant Analysis (HLDA). Since most identifiable accents are presented from the pronunciation of vowels rather than consonants \cite{Suzuki:2009}, multiple vowel-specific GMMs were computed with features of the vowel components, extracted either from phoneme alignment (in system development) or phoneme recognition (in system test). Compared with the baseline with pure acoustic information, the improved 7-way classification system increases accuracy from $42\%$ to $54\%$, using only up to 20 seconds speech data.   

This work was initiated during the author's internship at Interactive Intelligence (ININ) \cite{ge2013book}, and the algorithm and experiments were later refined for better accuracy and efficiency \cite{ge2015accent}. This paper reviews the major components of the accent classification system, with highlights on the recent improvements in feature generation and constructing baseline and improved classifiers. The remaining paper is organized as follows: Sec. \ref{sec:data_feature} introduces the database used in this work and a process of feature optimization and dimension reduction. In Sec. \ref{sec:vowel}, the main concept of creating accent-adapted features based on phonetic vowels is demonstrated. Then, the baseline classifier and improved version with vowel extraction are described in Sec. \ref{sec:gmm-ubm}, followed by the results, conclusion and future work in Sec. \ref{sec:results}.

\section{Data and Feature Preparation}
\label{sec:data_feature}

\begin{figure}[htb]
  \centering
 	\includegraphics[scale=0.5]{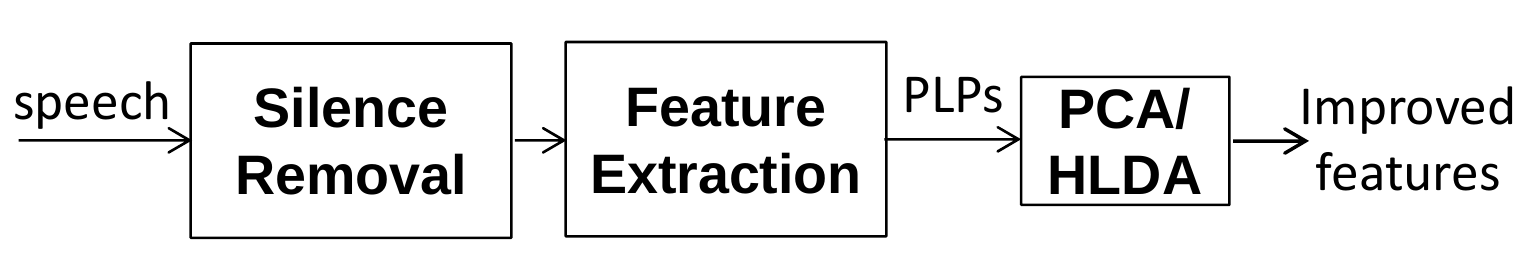}
    \caption{Process of data and feature preparation \label{fig:data_feature}}
\end{figure}

Preprocessing such as data and feature preparation, significantly impact the performance of classification. In this section, we introduce the database used in this work, discuss the feature extraction including normalization and Gaussianization, and feature optimization and dimension reduction with PCA and HLDA. The whole process is illustrated in \figurename \ref{fig:data_feature}. 

\subsection{Database}
\label{subsec:data}

The Foreign Accented English (FAE) corpus from Linguistic Data Consortium (LDC) with catalog number LDC2007S08 is used in this work. It is one of the most comprehensive accented English speech database currently available, which contains 4925 sentences of 23 types of accents, with 20 second duration on average.

Accents are divided into 7 major categories based on their relationships shown in Table  \ref{tab:fae_group} and one accent from each group is selected for developing a 7-way accent classifier.   
\begin{table}[htb]
%\myfigureshrinker
\centering
%\small
\caption{Type of accents in FAE corpus}
\label{tab:fae_group}
\renewcommand{\tabcolsep}{5pt}
\begin{tabular}{@{} cll @{}} \toprule
Group & Accents & Abbreviations \\\midrule
1 & \textbf{Arabic} & \textbf{AR} \\
2 & \textbf{French}, Indonesian, Malay, Swahili & \textbf{FR}, IN, MY, SW \\
3 & Cantonese, Japanese, Korean,  & CA, JA, KO, \\
  & \textbf{Mandarin}, Vietnamese & \textbf{MA}, VI \\
4 & \textbf{Brazilian Portuguese}, Italian, & \textbf{BP}, IT, \\ 
  & Iberian Portuguese, Spanish & PP, SP \\
5 & Czech, Hungarian, Polish, \textbf{Russian} & CZ, HU, PO, \textbf{RU} \\
6 & Farsi, \textbf{German}, Swedish & FA, \textbf{GE}, SD \\
7 & \textbf{Hindi}, Tamil & \textbf{HI}, TA \\
\bottomrule
\end{tabular}
%\myfigureshrinker
\end{table}
Table \ref{tab:fae_summary} summarizes some statistics in each of these accent groups, such as 1) the number of utterances; 2) their proportion in the entire FAE corpus, 3) the total durations before and after silence removal, and 4) their corresponding compression ratio ($\textrm{Dir}._{2}/\textrm{Dir}._{1}$). Since there is no transcription comes along with the speech in FAE, we also transcribed the audio data of these 7 major accents, in order to perform vowel extraction from speech using phoneme alignment later. The selected partial dataset of FAE were randomly divided into training, development and testing with ratio $70:15:15$.

\begin{table}[tb]
%\myfigureshrinker
\centering
%\small
\caption{Summary of selected accents in FAE corpus}
\label{tab:fae_summary}
\renewcommand{\tabcolsep}{2.5pt}
\begin{tabular}{@{} cccccc @{}} \toprule
Accents & No. of & Proportion & Total & Total & Compression \\
(Abbr.) & utterances & (\%) & Dur.$_1$ & Dur.$_2$ & ratio (\%) \\ \midrule
AR & 112 & 2.27 & 0:34:32 & 0:29:11 & 84.51 \\
BP & 459 & 9.32 & 2:34:24 & 2:09:58 & 84.18 \\
FR & 284 & 5.77 & 1:31:05 & 1:18:44 & 86.44 \\
GE & 325 & 6.60 & 1:36:04 & 1:22:18 & 85.67 \\
HI & 348 & 7.07 & 1:56:10 & 1:36:31 & 83.08 \\
MA & 282 & 5.73 & 1:30:37 & 1:16:06 & 84.98 \\
RU & 236 & 4.79 & 1:11:13 & 0:59:54 & 84.11 \\\midrule
Average & 292 & 5.93 & 1:33:26 & 1:18:57 & 84.57 \\
\bottomrule
\multicolumn{6}{p{0.85\linewidth}}{Dur.$_1$ and Dur.$_2$ are the duration before and after silence removal}
\end{tabular}
%\myfigureshrinker
\end{table}

\subsection{Silence Removal}
\label{subsec:silence_removal}

In practice, only the high signal-to-noise ratio (SNR) regions of the waveform are retained for classification. Therefore, silence removal or so-called voice activity detection (VAD) is often performed before feature extraction. Here we use the method described in  \cite{giannakopoulos2009method}, which detects the silence by thresholding on the short-time energy rate and spectral centroids of the speech. One can also use either Auditory Toolbox \cite{slaney1998auditory} or Voicebox \cite{brookes1997voicebox} for the same purpose.

Given $s_i(n), n \in [1,N]$ as the audio samples in the $i^\textrm{th}$ frame, its short-time energy rate $e_i$, can be formulated as
\begin{equation}
e_i = \frac{1}{N}\sum_{n=1}^{N}|s_{i}(n)|^{2} ,
\end{equation}  
where $N$ is the number of samples in one frame. The spectral centroid can be defined as
\begin{equation}
c_{i} = \frac{\sum_{k=1}^{K}(k+1)S_{i}(k)}{\sum_{k=1}^{K}S_{i}(k)} ,
\end{equation}
where $S_{i}(k), k \in [1,K]$ is the Discrete Fourier Transform (DFT) coefficients of $s_{i}$. $e_i$ is used to discriminate silence with environmental noise, while $c_i$ is used to remove non-speech noise, such as coughing, because of its lower energy concerntration in the spectrum, relative to that of normal human speech. 

\begin{figure}[tb]
  \centering
 	\includegraphics[scale=0.4]{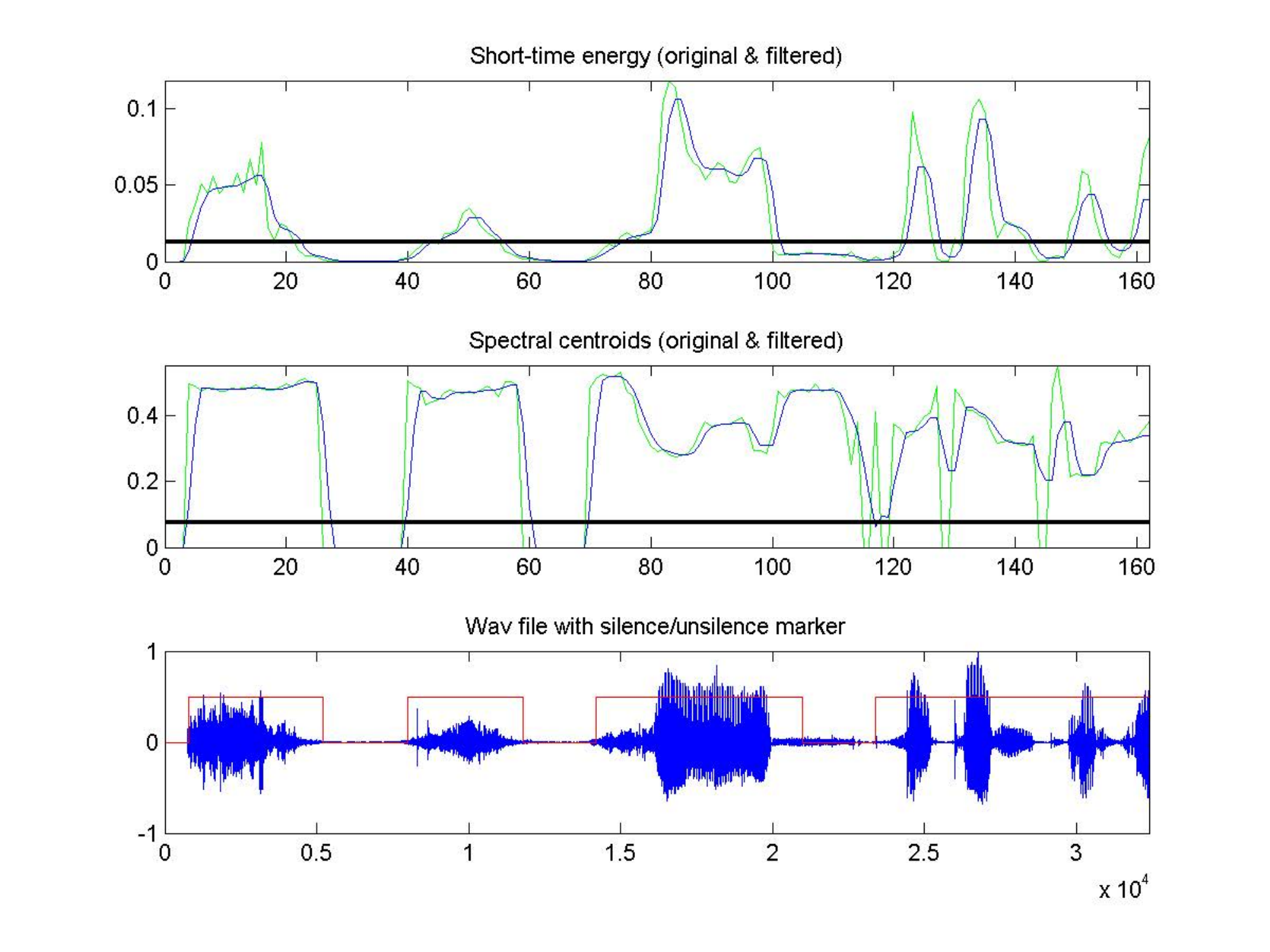}
    \caption{Example of silence removal using short-time energy rate and spectral centroids (FAR00035.wav in FAE) \label{fig:silence_removal_example}}
%\myfigureshrinker
\end{figure}

\figurename \ref{fig:silence_removal_example} shows an example of silence removal with both measurements of short-time energy rate and spectral centroids on data file FAR00035.wav in FAE corpus with Arabic (AE) accents. It is considered to be silence if either of these 2 measurements is lower than its threshold. As shown in Table \ref{tab:fae_summary}, the total duration of recording for each type of accents were reduced after silence removal.   

\subsection{Feature Extraction and Optimization}
\label{subsec:feat_gen}

After silence removal, the data of the selected accents were then transformed to 39-dimensional PLP windowed feature frames with $10$ millisecond each, using the method in \cite{plp}. Feature Mean and Variance Normalization (MVN) and short-term Gaussianization (a.k.a feature warping) were applied afterwards using the method in \cite{sadjadi2013msr}. The latter warps the distribution of the feature to a standard normal distribution to mitigate the effects of locally linear channel mismatch. This is specially useful because the features distribution here can be modeled by Gaussians \cite{pelecanos2001feature}.  

The normalized and wrapped features were further improved by PCA and HLDA for dimension reduction and optimization. PCA is commonly used for dimension reduction, which preserves the data dimensions with larger variations in the eigenspace. It has been applied to many applications, such as face recognition \cite{turk1991face} and speech evaluation \cite{ge2011pca}, etc. It also helps to regularize the data and avoid over-fitting in HLDA which is performed afterwards \cite{yang2003can}. Here the feature dimension is reduced from 39 to 30 after applying PCA. 

Compared with PCA, LDA reduces dimensions by mapping data into a subspace while maximizing the discriminative information. It overcomes the weakness  of PCA, when the discriminative information is actually in the dimension with less variation. It has been also applied to many problems, such as face recognition \cite{lu2003face} and speaker recognition \cite{ge2012pca}. Assume there are $K = \sum_{s=1}^{S}K_{s}$ number of $M$-dimensional data vectors $\bm{x}_k$ in $S$ classes, where $K_{s}$ is the number of vectors in class $s \in [1,S]$. Let the global mean $\bm{\Phi}$ over all classes be $\frac{1}{K}\sum_{k}^{K}\bm{x}_k$ and the local mean $\bm{\Phi}_s$ for each class $s$ be $\frac{1}{K_s}\sum_{\bm{x}_{k} \in s}\bm{x}_{k}$, the between-class scatter $S_{B}$ and within-class scatter $S_{W}$ can be defined as 
\begin{eqnarray} \label{eq:bcs}
	S_B &=& \frac{1}{K}\sum^{K}_{k=1}(\bm{x}_k-\bm{\Phi})(\bm{x}_k-\bm{\Phi})^{T} \ \textrm{or} \nonumber \\
	S_B &=& \frac{1}{S}\sum^S_{s=1}(\bm{\Phi}_s-\bm{\Phi})(\bm{\Phi}_s-\bm{\Phi})^T, 
\end{eqnarray}
and
\begin{eqnarray} \label{eq:wcs}
	S_W &=& \frac{1}{S}\sum^S_{s=1}\sum_{\bm{x}_k \in s}(\bm{x}_k-\bm{\Phi}_s)(\bm{x}_k-\bm{\Phi}_s)^{T} \ \textrm{or} \nonumber \\
	S_W &=& \frac{1}{S}\sum^S_{s=1}\frac{1}{K_s}\sum_{\bm{x}_k \in s}(\bm{x}_k-\bm{\Phi}_s)(\bm{x}_k-\bm{\Phi}_s)^T.
\end{eqnarray}
The first definitions of Eq. (\ref{eq:bcs}) and Eq. (\ref{eq:wcs}) consider the class weights, i.e. the sizes of each class $s$, while the second does not. The first definitions are consistent with the LDA definition used in Kumar's HLDA work \cite{Kumar:1997investigation} and are used in this work. However, the second definitions of both formulas are also provided for completeness. $S_{W}$ is likely to be singular if there is not enough data in that class, however, by applying PCA first, this problem can be significantly alleviated. 

Define $\bm{w} $ as a direction in the underlying $W$ to be transformed to, $\bm{w}^T S_B \bm{w}$ and $\bm{w}^T S_W \bm{w}$ are the projections of $S_B$ and $S_W$ onto $\bm{w}$ and searching for directions $w$ for the best class discrimination is equivalent to maximizing the ratio of $(\bm{w}^T S_B \bm{w}) / ({\bm{w}^T S_W \bm{w}})$ subject to $\bm{w}^T S_W \bm{w} = 1$. The latter is called the Fisher Discriminant function and can be converted by Lagrange multipliers and solved by eigen-decomposition of $S^{-1}_W S_B$. By selecting eigenvectors associated with the most significant $m$ eigenvalues of $S^{-1}_W S_B$, one can map the original $M$-dimensional data into a $m$-dimensional subspace for discriminative feature reduction.

LDA is derived with the assumption that features in various dimensions have the same variance, but in real scenario, there are examples illustrating LDA may transform data into a sub-optimal space when the dimension variances are different. Hetroscedastic LDA (HLDA) is a generalization of LDA using Maximum Likelihood Estimation (MLE) on Gaussian ditributions, which removes this assumption. An improved version of Kumar's HLDA algorithm with more flexibility and higher efficiency was developed in MATLAB and used in this work \cite{ge2013mispronunciation}. The context size $C$ is set to 1 and the feature dimension is further reduced from 30 to 20 after applying HLDA.  

\section{Phonetic Vowel Representation}
\label{sec:vowel}

Minematsu et al. \cite{Minematsu:2004} and Suzuki et al. \cite{Suzuki:2009} demonstrated that, for a particular speaker, the location of 5 fundamental vowels in the feature space of a target language (such as English in this work), is relatively consistent. Therefore, they can be extracted as accent-adapted features and used for identifying accent of that speaker.

\begin{figure}[htb]
  \centering
 	\includegraphics[scale=0.35]{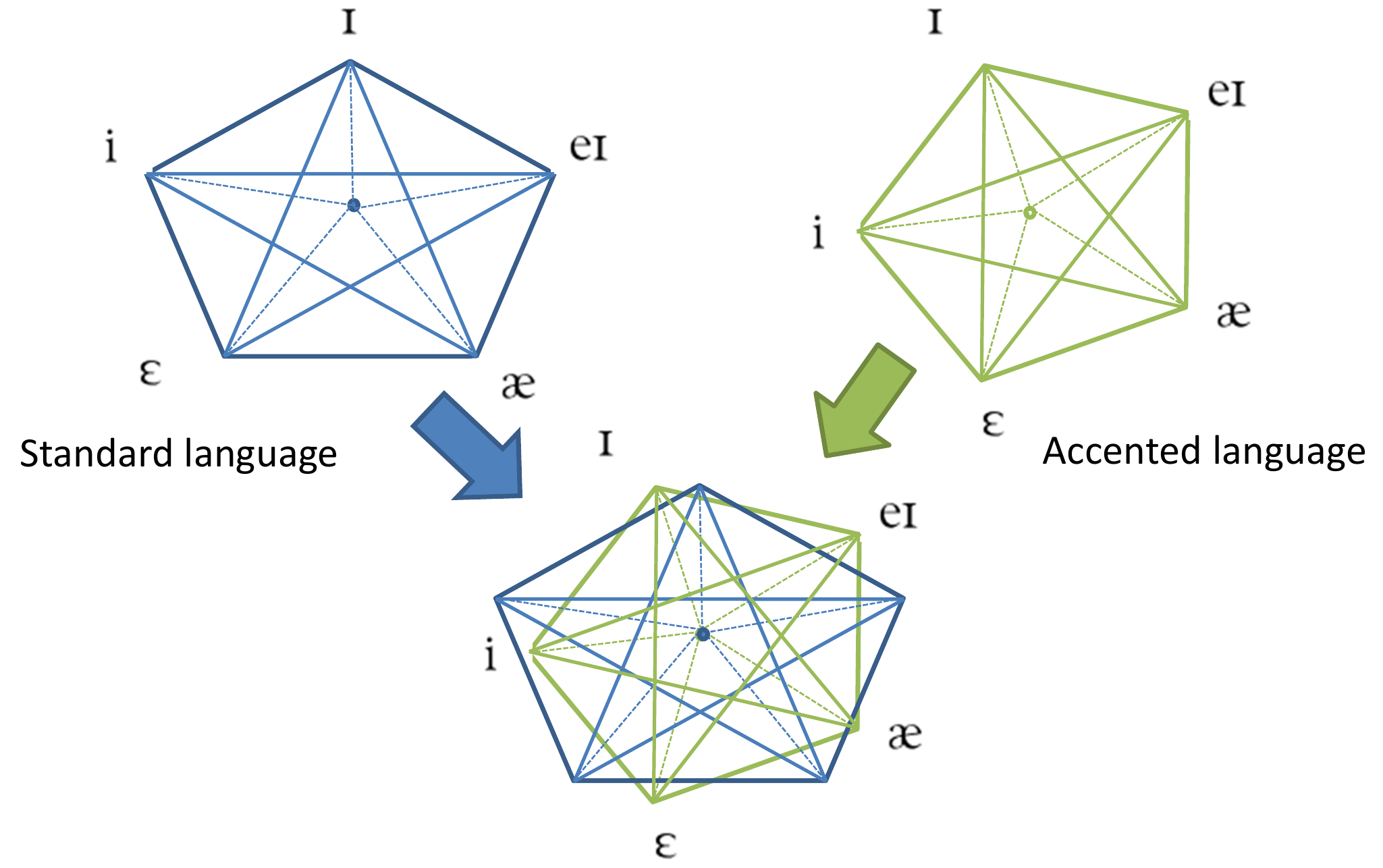}
    \caption{Comparison of 5 vowels locations in standard and accented language\label{fig:5vowels}}
\end{figure}

\figurename \ref{fig:5vowels} is a simple demonstration of 5 vowels from both accented and non-accented (standard) languages in the reduced 2-dimensional feature space \cite{Minematsu:2004}. The center in each pentagon is the weighted average of five vowels based on their positions in feature space and frequency of appearance in the corpus. By matching the center of the pentagon of the standard and the accented language into the overlapped pentagon in the bottom of Figure \ref{fig:5vowels}, the Bhattacharyya distances \cite{Bhattacharyya:1946} between each pair of corresponding vowels and their angles can be computed and stored in a vector. This vector $V_i$ represents the difference from the accented language $L_i$ to the standard one $L$.

To classify the test speech into one of the accent categories $L_1, L_2, \ldots, L_N$, where $N$ is the number of accent categories, the difference from $V_j$ to $V_i, i \in [1,N]$ and $V$ (category of standard language) are computed, compared and classified to the nearest category of accent. 

\subsection{Phoneme Alignment in System Development}
\label{subsec:phoneme_alginment}

\begin{figure*}[!htb]
  \centering
  \includegraphics[scale=0.55]{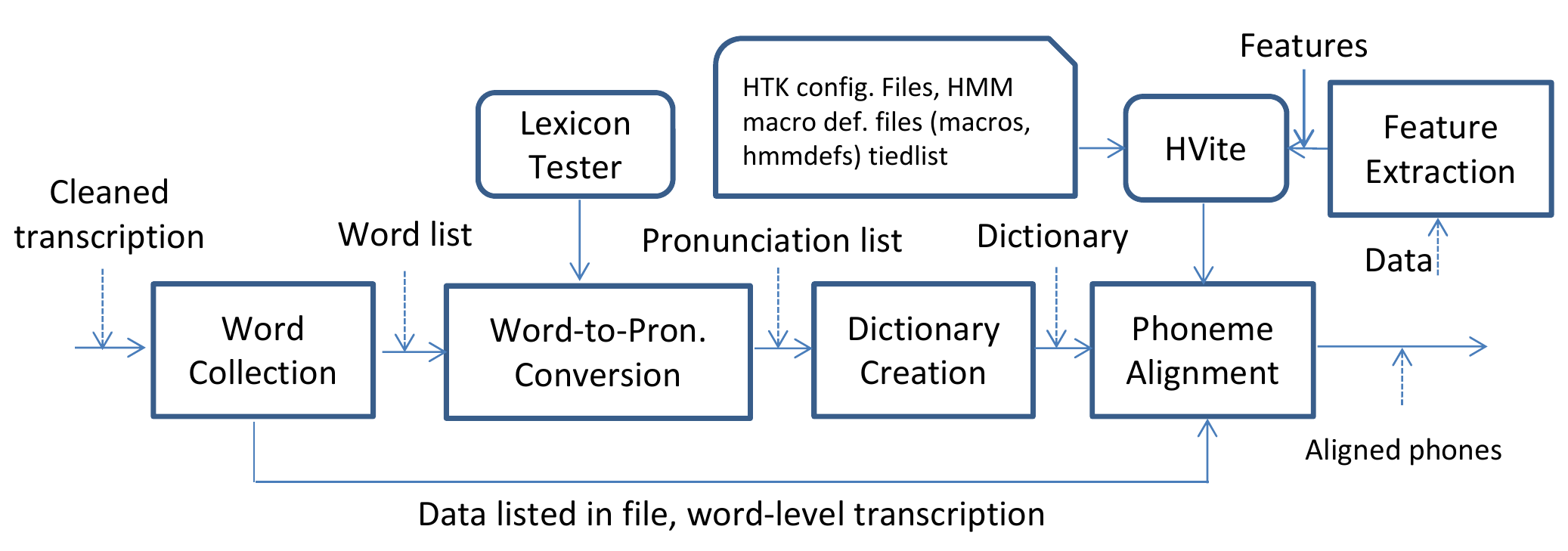}               
  \caption{Dictionary preparation and phoneme alignment for FAE corpus}
  \label{fig:dppa}
\end{figure*}

In order to extract vowels from speech data, phoneme alignment was performed during system development, with the in-house transcriptions of partial FAE corpus covering the 7 accents, each of which is from one major type of the accent groups. We prepared the dictionary needed for phoneme alignment using HVite in HTK \cite{young2006htk}, through a procedure including transcription cleaning, word collection, word-to-pronunciation conversion, etc. Figure \ref{fig:dppa} demonstrates the process of dictionary preparation and phoneme alignment for FAE corpus.   
The dictionary file is a list of pairs of words and pronunciations in HTK format, which can be obtained through the process of word collection, word-to-pronunciation conversion with ININ Lexicon Tester and HTK dictionary file creation. In Phoneme alignment, the HTK configuration file, HMM model definition and tired list were all trained using Fisher corpus.

\subsection{Phoneme Recognition in System Test}
\label{subsec:phoneme_recognition}

During system test, there is no transcription available. To find features corresponding to vowels, phoneme recognition was performed on the test accented speech using HTK. Since the recognition cannot be perfect, only a subset of recognized vowels with level of confidence score higher than a threshold based on the $n$-gram log likelihood were used. This threshold was predefined with the training and development data. 

\section{GMM-UBM Framework for Accent Classification}
\label{sec:gmm-ubm}

The Gaussian Mixture Model-Universal Background Model (GMM-UBM) framework has been sussessfully applied to speech verification and classification systems \cite{reynolds2000speaker}. The accent classification algorithm developed in this paper treats accents as speakers, and models the attributes of accents using GMM-UBM, which is similar to modeling the attributes of speakers in speaker classification systems. Modern speaker classification systems train a general Gaussian Mixture Model (GMM) with data from all speakers, so-called Universal Background Model (UBM) and then generate individual GMMs for each speaker by adapting UBM with features from individual speakers. Subsec. \ref{subsec:ubm} and \ref{subsec:gmm} provide an outline of applying the similar framework to an accent classification problem.

\subsection{Universal Background Model (UBM)}
\label{subsec:ubm}

The Universal Background Model (UBM) is a general GMM trained with features from all types of accents. Given a GMM $\lambda = \{w_{i}, \bm{\mu}_{i}, \Sigma_{i}\}$, $i \in [1,N]$ and $N$ is the number of mixture components, the likelihood function for a feature frame $\bm{x}$ can be formulated as 
\begin{equation} \label{eq:gmm_p}
	p(\bm{x}|\lambda) = \sum^N_{i=1} w_{i} p_{i}(\bm{x}), 
\end{equation}
where
\begin{equation} \label{eq:gmm_b}
p_{i}(\bm{x}) = \frac{1}{(2\pi)^{M/2}\vert\Sigma_i\vert^{1/2}} \mathrm{exp}\lbrace - \frac{1}{2}(\bm{x}-\bm{\mu}_i)^T\Sigma^{-1}_i(\bm{x}-\bm{\mu}_i)\rbrace.
\end{equation}
The parameters of GMM $\lambda$, including weight $w_{i}$, $\bm{\mu}_{i}$ and covariance matrix $\Sigma_{i}$ can be optimized by Expectation-Maximization (EM) algorithm \cite{dempster1977maximum}. Here $\Sigma_{i}$ is restricted to be diagonal. Usually the feature vectors are assumed independent, so the log-likelihood of a GMM $\lambda$ for a sequence of $K$ feature vectors, $X=\{\bm{x}_{1},\bm{x}_{2}, \ldots, \bm{x}_{K}\}$ is computed as
\begin{equation} \label{eq:loglik}
	\log p(X|\lambda) = \sum_{m=1}^{K} \log p(\bm{x}_{k}|\lambda).
\end{equation}

\subsection{Adaptation of Accent Model}
\label{subsec:gmm}

In the GMM-UBM system, we derive the individual accent GMMs by adapting the parameters of the UBM $\lambda_\mathrm{UBM}$ using the training speech $X=\{\bm{x}_{1},\bm{x}_{2}, \ldots, \bm{x}_{K}\}$ of each accents and a form of Bayesian adaptation. The adaptation is a two step process and the first step is identical to the E-step of the EM algorithm, where we determine the probabilistic alignment of $X$ into the UBM mixtures. That is for $i^{\textrm{th}}$ component of the UBM, we compute
\begin{equation} \label{eq:prob_align}
\mathrm{Pr}(i|\bm{x}_{k}) = \frac{w_{i}p_{i}(\bm{x}_k)}{\sum_{n=1}^{N}w_{n}p_{n}(\bm{x}_k)} .
\end{equation}
Then, the weight, mean and variance can be computed by
\begin{equation}
n_{i} = \sum_{k=1}^{K} \mathrm{Pr}(i|\bm{x}_{k}) ,
\end{equation}
\begin{equation}
E_{i}(\bm{x}) = \frac{1}{n_{i}} \sum_{k=1}^{K} \mathrm{Pr}(i|\bm{x}_{k})\bm{x}_{k} , 
\end{equation}
\begin{equation}
E_{i}(\bm{x}^{2}) = \frac{1}{n_{i}} \sum_{k=1}^{K} \mathrm{Pr}(i|\bm{x}_k)\bm{x}_{k}^{2},
\end{equation}
where $\bm{x}^{2}$ is shorthand for diag($\bm{x}\bm{x}^{T}$). Finally, these new sufficient statistics from the training data are sued to update UBM for mixture $i$, to create the adapted parameters for mixture $i$ in the accent GMMs, with the equations:
\begin{equation}
\hat{w}_{i} = [\alpha_{i}^{w}n_{i}/T + (1-\alpha_{i}^{w})w_{i}] \gamma
\end{equation}
\begin{equation}
\hat{\bm{\mu}_{i}} = \alpha_{i}^{m}E_{i}(\bm{x}) + (1-\alpha_{i}^{m})\mu_{i}
\end{equation}
\begin{equation}
\hat{\sigma}_{i}^{2} = \alpha_{i}^{v}E_{i}(\bm{x}^2) + (1-\alpha_{i}^{v})(\sigma_{i}^{2}+ \mu_{i}^{2}) - \hat{\mu}_{i}^{2} ,
\end{equation}
where $\{\alpha_{i}^{w}, \alpha_{i}^{m}, \alpha_{i}^{v}\}$ are the adaptation coefficients for the weights, means and variances respectively, controlling the balance between old and new estimates. They can be derived from the Maximum \textit{a posteriori} (MAP) estimation equations for a GMM using constraints on the prior distribution described in \cite{gauvain1994maximum}. The scale factor $\gamma$ is computed over all adapted mixture weights to ensure they sum to 1.

\subsection{Baseline Classifier}
\label{subsec:baseline}

\begin{figure}[htb]
  \centering
 	\includegraphics[scale=0.5]{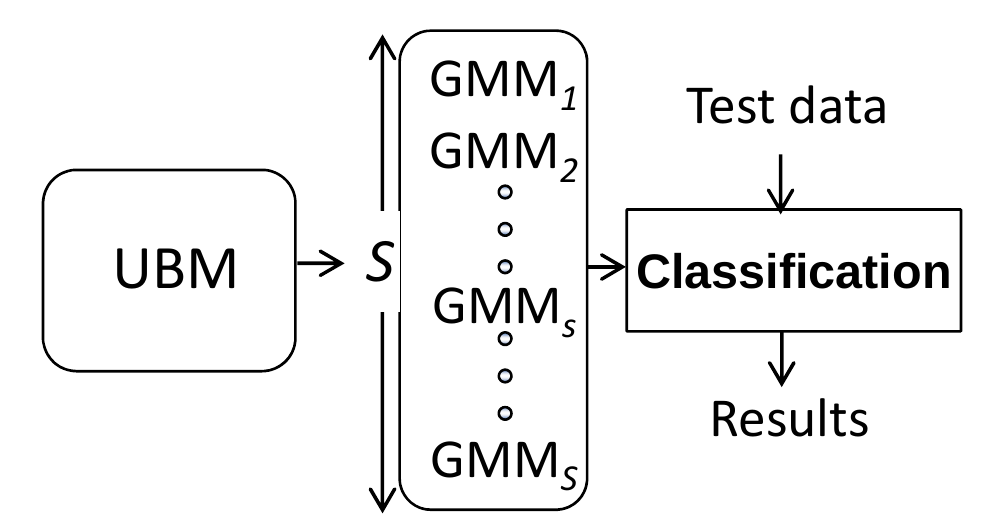}
    \caption{Baseline GMM classifiers adapted from UBM \label{fig:baseline}}
\end{figure}

After obtaining the adapted GMM parameter set $\lambda_s$ through GMM-UBM framework for accent class $s \in [1,S]$, the GMM-based classifier, which maximize \textit{a posteriori} probability for $K$ $M$-dimensional feature vectors $X$ ($M \times K$) can be formulated as:
\begin{eqnarray} \label{eq:gmmclassifier}
\small
	\hat{S} & = & \argmax_{s \in [1,S]}\mathrm{Pr}(\lambda_s|X) = \argmax_{s \in [1,S]}\frac{p(X|\lambda_s)\mathrm{\lambda_s}}{p(X)} \nonumber \\
			& \propto & \argmax_{s \in [1,S]}p(X|\lambda_s)  \nonumber \\
			& \propto & \argmax_{s \in [1,S]}\sum^{K}_{k=1}\mathrm{log}p(\bm{x}_{k}|\lambda_s).
\end{eqnarray}
The first equation is due to Bayes' rule. The first proportion is assuming $\mathrm{Pr}(\lambda_s) = 1/S$ and $p(X)$ is the same for all accent models. The second proportion uses logarithm and independence between input samples $\bm{x}_k$, $k \in [1,K]$, explained before in Eq. (\ref{eq:loglik}). Providing accent features for UBM adapation, \figurename \ref{fig:baseline} shows the diagram for baseline classification with accent GMM classifiers.

\subsection{Improved Classifier}
\label{subsec:improved}

\begin{figure}[htb]
  \centering
  	\includegraphics[scale=0.4]{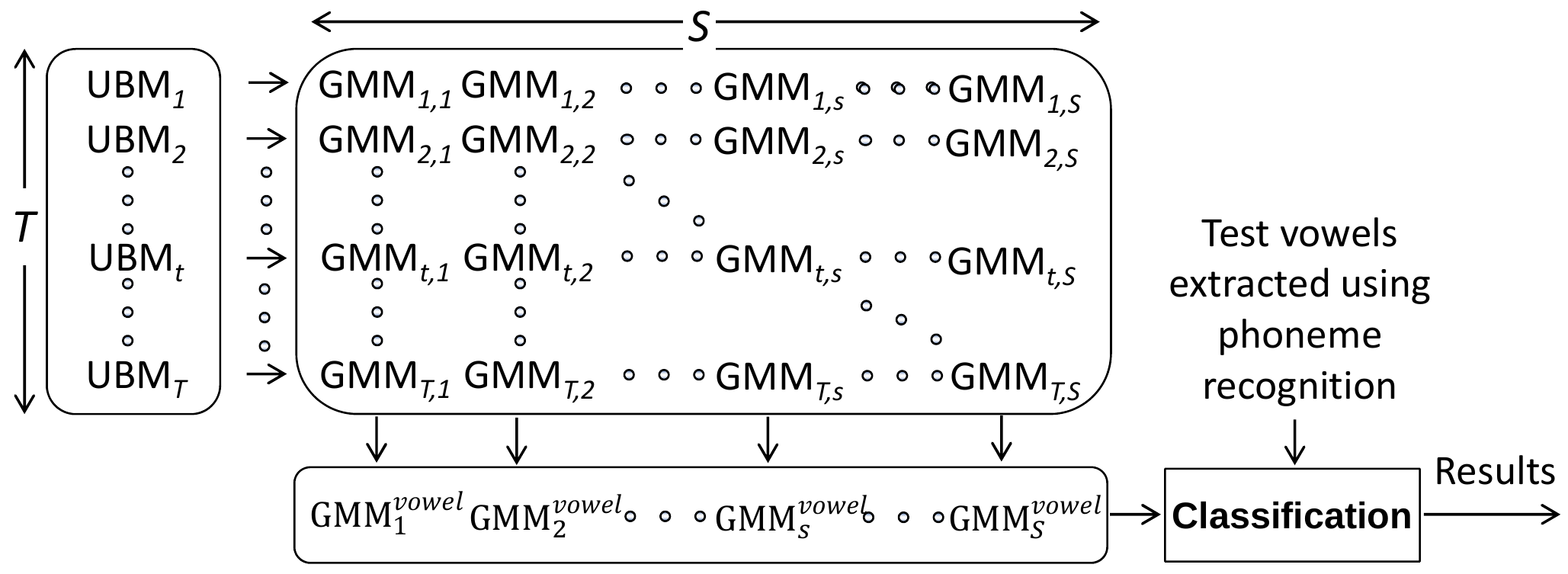}	
 	%\centerline{\hspace{2em}\includegraphics[scale=0.4]{gmm-ubm_vowel.pdf}}
    \caption{Improved GMM classifiers with vowel extraction \label{fig:improved}}
\end{figure}

The improvement from the baseline is mainly contributed from the vowel extraction. To construct the classifier with vowel representation, instead of directly measureing the shift of vowels from the standard speech to the accented one, the speech segments from the same vowel of various types of accents were concatenated and used to train vowel-specific UBMs. \figurename \ref{fig:improved} shows that each of the $T$ UBMs was then adapted to $S$ separated GMMs using data from various accent types.
Here $T$ is the number of vowels used in this work. Instead of using only the fundamental 5 vowels described in Sec. \ref{sec:vowel}, the same concept was generalized and all 15 vowels in Arpabet \cite{arpabet} were used, which are listed in Table \ref{tab:vowels}. $S$ is the number of selected accent types from FAE corpus (Table \ref{tab:fae_summary}), which is 7 in this work.
\begin{table}[htb]
\centering
\small
\caption{Vowels in Arpabet}
\label{tab:vowels}
\renewcommand{\tabcolsep}{2pt}
\renewcommand\baselinestretch{1}\selectfont
\begin{tabular}{@{} {l}*{9}{l} @{}} \toprule
\textbf{Vowel} & aa & ae & ah & ao & aw & ay & eh & er \\ 
\textbf{Example} & f\textbf{a}ther & f\textbf{a}st & s\textbf{u}n & h\textbf{o}t & h\textbf{ow} & m\textbf{y} & r\textbf{e}d & b\textbf{ir}d  \\\midrule
\textbf{Vowel} & ey & ih & iy & ow & oy & uh & uw & \\
\textbf{Example} & s\textbf{ay} & b\textbf{i}g & m\textbf{ee}t & sh\textbf{ow} & b\textbf{oy} & b\textbf{oo}k & f\textbf{oo}d \\\bottomrule
\end{tabular}
\end{table}
Given extracted features of $T$ types of vowels $[X^{(1)}, X^{(2)}, \ldots, X^{(t)}, \ldots, X^{(T)}]$ from accented test feature $X$, the improved GMM accent classifier as the combination of GMM classifiers of all vowels (shown at the bottom box in \figurename \ref{fig:improved}) can be formulated as 
\begin{eqnarray} \label{eq:vrclassifier}
	\hat{S} & = & \argmax_{s \in [1,S]}\sum_{t=1}^{T}w_{t}\mathrm{Pr}(\lambda_{s,t}|X^{(t)}) \nonumber \\
			& \propto & \argmax_{s \in [1,S]} \sum_{t=1}^{T}w_{t}\sum^{K(t)}_{k=1}\mathrm{log}p(\bm{x}_{k}^{(t)}|\lambda_{s,t}).
\end{eqnarray}
where $\lambda_{s,t}$ is the GMM for $s^\textrm{th}$ accent and $t^\textrm{th}$ type of vowels, and $w_{t}$ is the weight of the vowel-specific GMM classifier for $t^\textrm{th}$ vowel. Adding this additional layer on the GMM classifier is critical to find the vowel sets which preserve the accents and later shown to improve on classifying accents.

\begin{figure}[htb]
  \centering
  	\includegraphics[scale=0.6]{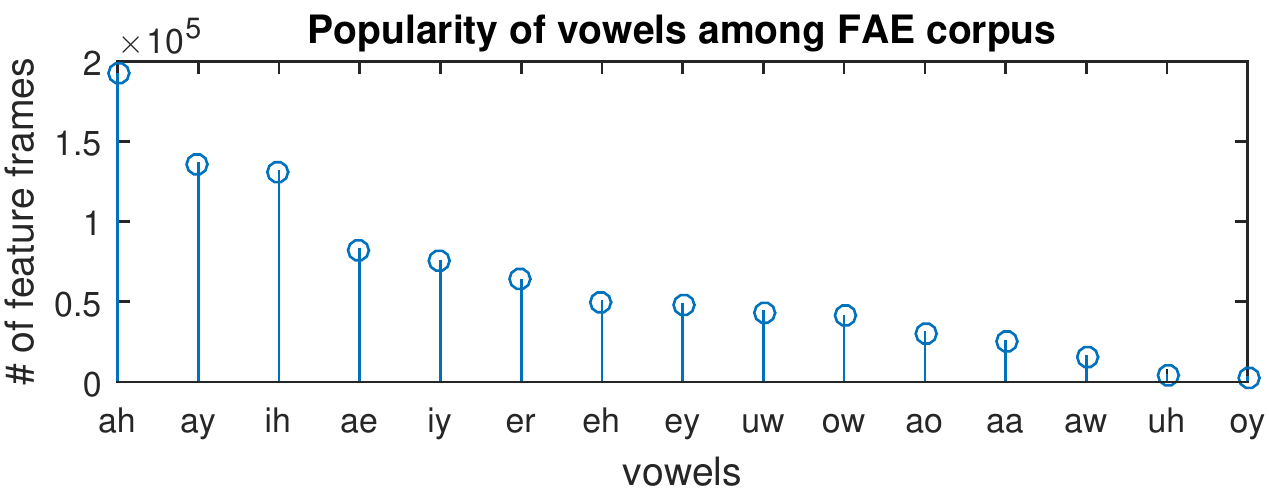}	
 	%\centerline{\hspace{2em}\includegraphics[scale=0.4]{gmm-ubm_vowel.pdf}}
    \caption{Popularity of vowels in the descending order of the corresponding feature frames in training data \label{fig:popularity}}
\end{figure}

\begin{figure}[htb]
  \centering
  	\includegraphics[scale=0.75]{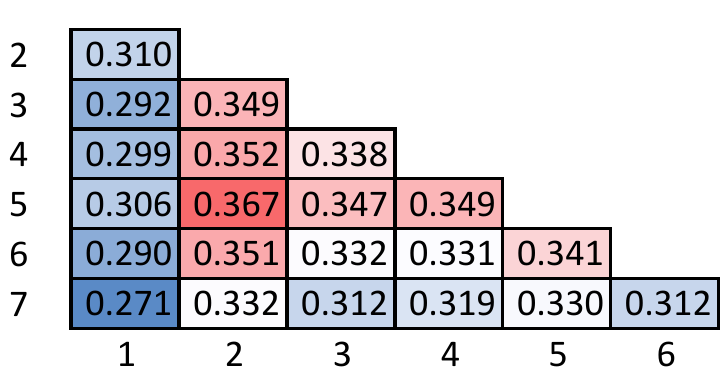}	
    \caption{Example of distance measurement among GMMs of same vowel ({\tt aa}) but 7 different accent types \label{fig:vowel_distribution_distance}}
\end{figure}

There are two factors considered in the vowel weight $w_t$ in Eq. (\ref{eq:vrclassifier}), which are 1) the popularity (proportion) of $t^\textrm{th}$ vowels in the whole vowel set $r_t$, and 2) the discriminativeness $d_t$, i.e. the difference in the distributions of the same vowels extracted from different accents. The first factor is based on the assumption GMMs trained with more data is more reliable than the ones with less data. \figurename \ref{fig:popularity} shows the popularity of vowels in descending order. It show the vowel {\tt ah} is much more popular (frequent) than the vowel {\tt oy} in the selected dataset.
The second factor is based on the assumption that vowels are more discriminative if the distributions of GMMs of the same vowel but different accents are far apart. For example, \figurename \ref{fig:vowel_distribution_distance} shows the Hellinger distances \cite{kristan2011multivariate} computed between any 2 GMM distributions of 7 different accent types for the vowel {\tt aa}. The discriminativeness factor $d_t$ for {\tt aa} is just the reciprocal of the mean of these distances from $\binom{7}{2}$ combinations (the smaller mean, the more weight).
Here we simply used $w_{t} = r_{t}d_{t}$ to compute the weights of vowels, which assumes both factors are equally important. 

\section{Results and Conclusion}
\label{sec:results}

The baseline classification was based on accent GMM classifier with 256 mixtures, adapted from UBM, using normalized and warped 39-dimensional PLPs. The features were then optimized using PCA and HLDA with context size $C=1$, and the dimension was reduced from 39 to 20. With the enhanced feature, the baseline accuracy was increased from $42.3\%$ to $47.9\%$. The main contribution of accuracy improvement was from the classifier of the combination of weighted vowels, which further increased the 7-way classification rate to $53.7\%$. Table \ref{tab:result_accentclassification} shows the performance of all these 3 experiments with various models and features.    
\begin{table}[htb]
\centering
%\footnotesize
\small
\caption{7-way accent classification under GMM-UMB framework with acoustic and phonetic features}
\label{tab:result_accentclassification}
\begin{tabular}{@{} llll @{}} \toprule
Model & $\mathbf{GMM}_{base}^{256}$ & $\mathbf{GMM}_{base}^{256}$ & $\mathbf{GMM}_{vowel}^{256}$ \\
Feature & $\mathbf{PLP}_{MVN}^{39}$ & $\mathbf{PCA/HLDA}_{C1}^{20}$ & $\mathbf{PCA/HLDA}_{C1}^{20}$ \\
Accuracy & 42.3\% & 47.9\% & 53.7\% \\\bottomrule
\end{tabular}
\end{table}

This work demonstrates that methods in speaker recognition can be used for accent classification. With several feature optimization techniques and phonetic vowel information, the accuracy obtained from accented speech as short as 20 seconds, is competitive compared with the state of the art in \cite{choueiter2008empirical}, \cite{omar2010novel} and \cite{macias2003acoustic}. In the future, more recent classification methods such as i-vector (eigenvoice component) \cite{matvejka2011full}, or neural network classifier \cite{ge2015sleep} can be explored, used or combined with the current methods. More data-driven techniques can be experimented, such as 1) training distinct UBMs for male and female accented speakers, 2) using tri-phone vowel set instead of the current mono-phone vowel set for more refined classification, 3) selecting a subset of vowels rather than using all 15 vowels in Arpabet by experiment for better classification results, etc.

\bibliographystyle{IEEEtran}
\bibliography{paper2}  
  
\end{document}